\documentstyle[12pt]{article}
\title{On one dimensional digamma and polygamma series related to the 
evaluation of Feynman diagrams}  %title modified too .............
\author{Mark W. Coffey\\
Department of Physics\\
Colorado School of Mines\\
Golden, CO  80401\\
(Received $\mbox{~~~~~~~~~~~~~~~~~~~~~~~~~~~~~~~2005}$)}
\date{January 2, 2005}
\begin{document}
\maketitle
%\vspace{.25cm}
\baselineskip=25 pt
\begin{abstract}

We consider summations over digamma and polygamma functions, often with
summands of the form $(\pm 1)^n\psi(n+p/q)/n^r$ and $(\pm 1)^n\psi^{(m)}
(n+p/q)/n^r$, where $m$, $p, q$, and $r$ are positive integers.  
We develop novel general integral representations and present explicit
examples.  Special cases of the sums reduce to known linear Euler sums.  
The sums of interest find application in quantum field theory, including
evaluation of Feynman amplitudes.

\end{abstract}
 
\vspace{.25cm}
\baselineskip=15pt
\centerline{\bf Key words and phrases}
\medskip 

\noindent
gamma function, digamma function, polygamma function, Riemann zeta function, 
Clausen function, Euler sums, harmonic numbers, Hurwitz zeta function,
hypergeometric function, dilogarithm function, trilogarithm function, 
polylogarithm function

%\vspace{.25cm}
%\vfill
%\centerline{\bf 2000 PACS numbers}
%TBS ...%02.30.-f, 06.30.Lz     %these are not specific 
\vfill
\centerline{\bf AMS Subject Classification codes}
33B, 33E, 11M
 
\baselineskip=25pt
\pagebreak
%\medskip
\centerline{\bf Introduction}
\medskip

The evaluation of Euler sums 
\cite{bailey,berndt,borwein,borwein95,bbk,coffey02,doelder,flajolet,freitas} 
%give bunch of other refs. here ...  give in alphabetic order here ... to do ...
is not only a beautiful subject in classical 
real and complex analysis, but has application in quantum field theory.
Specifically, in quantum electrodynamics, one evaluates Feynman amplitudes
from certain integrals \cite{qedref,nucl}.  %cite a Feynman book(s), e.g.
These often lead to Mellin transforms and Euler sums 
\cite{broadhurst,ogreid98,ogreid99,ogreid00,jamv}.  %cite then broadhurst paper(s), etc.
The problem of closed form evaluation of Euler sums has stimulated 
research in symbolic computation and integer relation detection techniques
\cite{bailey,jamv}.

Among many possible variations (e.g., \cite{bailey,rao}), various one
dimensional Euler sums may be written as
$$S_{r,p,q}=\sum_{n=1}^\infty {{[H_n^{(r)}]^p} \over n^q}, ~~~~
A_{r,p,q}=\sum_{n=1}^\infty (-1)^n {{[H_n^{(r)}]^p} \over n^q}, \eqno(1a)$$
$$S'_{r,p,q}=\sum_{n=1}^\infty {{[h_n^{(r)}]^p} \over n^q}, ~~~~
A'_{r,p,q}=\sum_{n=1}^\infty (-1)^n {{[h_n^{(r)}]^p} \over n^q}, \eqno(1b)$$
where the generalized harmonic numbers are given by
$$H_n^{(r)} \equiv \sum_{j=1}^n {1 \over j^r}, ~~~~H_n \equiv H_n^{(1)},
~~~~~~~~ h_n^{(r)} \equiv \sum_{j=1}^n {{(-1)}^{j+1} \over j^r}. \eqno(2)$$
A rich combination of elementary and special functions arises in the 
evaluation of Euler sums.  Special functions which typically appear are
the polylogarithms, gamma, digamma, and polygamma functions, zeta functions, 
Clausen functions, hypergeometric functions, tangent integral, and 
logarithmic-trigonometric integrals 
\cite{freitas,bailey,coffey02}.

%This paper is restricted to one dimensional Euler sums of a relatively
%simple sort that are very closely connected mainly with sums over the 
%polygamma functions $\psi^{(j)}$.  
This paper considers sums over digamma and polygamma functions at rational
argument, wherein special cases reduce to one dimensional linear Euler sums.
The tight connection between harmonic 
numbers and the digamma function is clear:  $H_n = \psi(n+1)-\psi(1) 
= \psi(n+1)+\gamma$, where $\gamma$ is the Euler constant.  
This extends to generalized harmonic numbers and the polygamma function,
$$H_n^{(r)}={{(-1)^{r-1}} \over {(r-1)!}}\left[\psi^{(r-1)}(n+1)-\psi^{(r-1)}
(1) \right ],$$
where $\psi^{(r-1)}(1)=(-1)^r(r-1)!\zeta(r)$ and $\zeta$ is the Riemann
zeta function \cite{edwards,karatsuba,riemann,titch}, so that
$$S_{r,1,q}={{(-1)^{r-1}} \over {(r-1)!}}\sum_{n=1}^\infty {{\psi^{(r-1)}
(n+1)} \over n^q} -(-1)^{r-1}\zeta(r)\zeta(q), ~~~~~~ r, q > 1.$$
Another set of numbers that often figures prominently in special cases of
our summations is
$$\gamma+\psi(n+1/2)=-2\ln 2+2\sum_{j=1}^n {1 \over {2j-1}}.  $$

Our results extend many of the previous 
sums calculated by de Doelder \cite{doelder}.  A significant physical
motivation is provided by the very recent work of Ogreid and Osland
\cite{ogreid98,ogreid99,ogreid00}, and we build upon their discussion.
These authors have calculated one and two dimensional Euler series 
associated with Feynman diagrams, invoking properties of the dilogarithm
and trilogarithm functions.
Our focus is not with polylogarithms \cite{lewin}, but to bring to bear 
especially further properties of the polygamma and Clausen functions.

%This paper is further restricted to linear Euler sums, those with
%$p=1$ in Eq. (1), although much of the technique extends to higher degree
%sums.  
Previously, we considered several nonlinear Euler sums, including
one closely related to $S_{1,4,2}$, $\sum_{n=1}^\infty H_n^4/(n+1)^2$
\cite{coffey02}.  We discussed the application of logarithmic-trigonometric
integral, Fourier series, and contour integral methods.  We conjecture this 
sum to have the value $(859/24)\zeta(6) + 3\zeta^2(3)$ \cite{bailey}, where 
%$\zeta(s)$ is the Riemann zeta function and 
$\zeta(6)=\pi^6/945$.

The sum $S_{1,1,q}$ has a closed form, and a history that seems to go back
to Euler \cite{rao}.  The companion alternating sums $A_{1,1,q}$ and
$A'_{1,1,q}$ can be written as a polynomial in zeta values when the weight 
$1+q$ is odd, while the alternating sum $S'_{1,1,q}$ has a closed form for
any weight \cite{rao,flajolet}.  In Ref. \cite{rao}, Sitaramachandra Rao 
formulated twelve classes of infinite series.  With his notation, 
$S_{1,1,q}=H_1(q)$ and $A_{1,1,q} = -H_7(q)$.  Among many useful results and 
consideration of a Ramanajun sum, this author gives the closed form for the 
sum $S'_{1,1,q}=H_4(q)$.

%LATER:  List all of the various 1D sum results contained herein ............
We first consider sums over the digamma function, containing
summands with $(\pm 1)^n \psi(n+p/q)/n^2$.  Such sums are amenable to
much simplification and several explicit alternative forms, owing to the 
properties of the Clausen function Cl$_2$.  This function is known at
rational multiples of $\pi$ and in fact is expressible in terms of the 
trigamma function at rational argument \cite{doelder84}.  
We next present similar results for summands with $(\pm 1)^n \psi(n+p/q)/n$,  
before proceeding to sums over $(\pm 1)^n \psi(kn)/n^p$.  For these more
general sums we construct new integral representations and illustrate
their reduction in a few special cases.  We then make an extension to
sums over the polygamma function.  The integral representations presented
here too subsume several previous results.  Before concluding, we offer
several variations on digamma and polygamma series with both integer and 
rational arguments and make a suggestion for the further study of linear
digamma and polygamma series with reciprocal binomial coefficients. 
We have relegated to an Appendix a quick estimation of many of the
digamma and polygamma sums of interest.

%someplace:  to go over my approximate comparison integrals etc. for the
%corresponding Euler sums ..................

\centerline{\bf Sums over $(\pm 1)^n \psi(n+p/q)/n^2$}
\medskip

We first present a general result for sums over the digamma function at
rational argument.  For integers $q \geq 1$ we have
{\newline \bf Proposition}
$$\sum_{n=1}^\infty {1 \over n^2}\sum_{r=0}^{q-1} \left[\gamma+\psi\left(n+
{r \over q}\right)\right ]=q\left[\left({q^2 \over 2}+{1 \over {2q}}\right)
\zeta(3)+\pi \sum_{j=1}^{q-1} \mbox{Cl}_2\left({{2\pi j} \over q}\right)
\right] - q \ln q \zeta(2), \eqno(3a)$$
$$\sum_{n=1}^\infty {{(-1)^n} \over n^2}\sum_{r=0}^{q-1} \left[\gamma+\psi
\left(n+{r \over q}\right)\right ]=q\left[\left({q^2 \over 2}-{3 \over {8q}}
\right) \zeta(3)+\pi \sum_{j=1}^{q-1} \mbox{Cl}_2\left({{2\pi j} \over q}
+{\pi \over q} \right) \right] + {q \over 2} \ln q \zeta(2), \eqno(3b)$$
where the Clausen integral is given by \cite{nbs,doelder84,grosjean}
$$\mbox{Cl}_2(\theta)=-\int_0^\theta \ln[2\sin(t/2)] dt=\sum_{k=1}^\infty
{{\sin k\theta} \over k^2}$$
$$=\int_0^1 \tan^{-1}\left({{\rho \sin \theta} \over
{1-\rho \cos \theta}}\right ) {{d\rho} \over \rho}=-\sin \theta\int_0^1
{{\ln \rho} \over {\rho^2-2\rho \cos \theta+1}} d\rho.  \eqno(4)$$

For proof, we start with the multiplication formula for the digamma 
function \cite{andrews,grad}, 
$$\sum_{r=0}^{q-1} \psi\left(b+{r \over q}\right)=q \psi(bq)-q \ln q.
\eqno(5)$$
We apply this formula with real $b$ put to integer $n$, multiply by
$(\pm 1)^n/n^2$ on both sides, and sum from $n=1$ to $\infty$, resulting in
$$\sum_{n=1}^\infty {1 \over n^2}\sum_{r=0}^{q-1} \left[\gamma+\psi\left(n+
{r \over q}\right)\right ]=q\sum_{n=1}^\infty {1 \over n^2}\left[\gamma
+\psi\left(qn\right)\right ]-q\ln q\zeta(2), \eqno(6a)$$
and
$$\sum_{n=1}^\infty {{(-1)^n} \over n^2}\sum_{r=0}^{q-1} \left[\gamma+\psi
\left(n+{r \over q}\right)\right ]=q\sum_{n=1}^\infty {{(-1)^n} \over n^2}
\left[\gamma+\psi\left(qn\right)\right ]+{q \over 2}\ln q\zeta(2). \eqno(6b)$$
For Eq. (6b), we used the infinite series for the alternating zeta function,
Li$_n(-1)=(2^{1-n}-1)\zeta(n)$, where Li$_q(z)=\sum_{n=1}^\infty z^n/n^q$ is
the polylogarithm.  By applying a Corollary of Ref. \cite{ogreid98}, we 
obtain the Proposition.

The case of $q=1$ is highly degenerate in Eq. (3).  Then both the Cl$_2$
and logarithm terms vanish, resulting in simply $\zeta(3)$ for (6a) and
$\zeta(3)/8$ for (6b).  We work through a series of additional special
cases, due to the applications interest, in order to make connections
with the literature, and because the Clausen functions may be much simplified.

For $q=2$ in Eq. (3), we obtain from the previous $q=1$ case
$$\sum_{n=1}^\infty {1 \over n^2}\left[\gamma+\psi\left(n+{1 \over 2}\right)
\right ]={7 \over 2}\zeta(3)-2 \ln 2 \zeta(2), \eqno(7a)$$
$$\sum_{n=1}^\infty {{(-1)^n} \over n^2}\left[\gamma+\psi\left(n+{1 \over 2}
\right)\right ]={7 \over 2}\zeta(3)-2\pi G+ \ln 2 \zeta(2), \eqno(7b)$$
where \cite{grosjean} %add other ref. here?
$$G = \mbox{Cl}_2\left({\pi \over 2}\right) = \sum_{m=0}^\infty {{(-1)^m}
\over {(2m+1)^2}} $$
$$={1 \over {16}}[\psi'(1/4)-\psi'(3/4)] \simeq 0.915965594177219015, 
\eqno(8)$$ 
is Catalan's constant, and $\psi'$ is the trigamma function.
Since $-\psi(1/2)=\gamma+2\ln 2$, Eqs. (7) agree
with results obtained by de Doelder \cite{doelder} by way of logarithmic
integrals.  Equations (7) also result from the direct combination of Eqs. 
(B.1) and (B.3) of Ref. \cite{ogreid98}.

We can just as well apply the method of proof of Eq. (3) with $b \to 2n$
in Eq. (5) before summing over $n$.  This gives, for instance,
$$\sum_{n=1}^\infty {1 \over n^2}\left[\gamma+\psi\left(2n+{1 \over 2}\right)
\right ]=14\zeta(3)-4\pi G -2 \ln 2 \zeta(2). \eqno(9a)$$
This equation also follows from the direct combination of Eqs. (B.3) and
(B.7) of Ref. \cite{ogreid98}.  The corresponding alternating sum makes
use of a number of the properties of the Clausen function, so that we
record it as
%see p. 9 of my notes of 10/09/04 pm
{\newline \bf Lemma}
$$\sum_{n=1}^\infty {{(-1)^n} \over n^2}\left[\gamma+\psi\left(2n+{1 \over 2}
\right)\right ]=14\zeta(3)+\zeta(2)\ln 2+2\pi G -8\pi \mbox{Cl}_2(\pi/4), 
\eqno(9b)$$
where \cite{grosjean}
$$\mbox{Cl}_2(\pi/4)={1 \over {32}}[\sqrt{2}\psi'(1/8)-2(\sqrt{2}+1)\pi^2
-8(2\sqrt{2}-1)G], \eqno(10)$$
%and $\psi'$ is the trigamma function.--mentioned above.
In obtaining Eq. (9b), the expression
Cl$_2(3\pi/4)+2$Cl$_2(5\pi/4)+3$Cl$_2(7\pi/4)$ has been reduced by
using the symmetry Cl$_2(\pi+\theta)=-$Cl$_2(\pi-\theta)$ and the
duplication formula satisfied by the Clausen function.  There results
Cl$_2(5\pi/4)=-\mbox{Cl}_2(3\pi/4)$, Cl$_2(7\pi/4)=-\mbox{Cl}_2(\pi/4)$, 
and Cl$_2(3\pi/4) = \mbox{Cl}_2(\pi/4)-G/2$.

As a follow on to Eqs. (9) we have
$$\sum_{n=1}^\infty {1 \over n^2}\left[\gamma+\psi\left(3n+{1 \over 2}\right)
\right ]={{63} \over 2}\zeta(3)-10\pi \mbox{Cl}_2(\pi/3)-2\ln 2 \zeta(2), 
\eqno(11)$$
where \cite{grosjean}
$$\mbox{Cl}_2(\pi/3)={1 \over {2\sqrt{3}}}[\psi'(1/3)-2\pi^2/3].  \eqno(12)$$
For the corresponding alternating sum we find
{\newline \bf Lemma}
$$\sum_{n=1}^\infty {{(-1)^n} \over n^2}\left[\gamma+\psi\left(3n+{1 \over 2}
\right)\right ]={{63} \over 2}\zeta(3)+3\pi \mbox{Cl}_2(\pi/3)+\ln 2 \zeta(2)
-{{26} \over 3}\pi G -8\pi \mbox{Cl}_2(\pi/6)$$
$$={{63} \over 2}\zeta(3)+\zeta(2)\ln2-14\pi G. \eqno(13)$$
In obtaining Eq. (13), relations such as Cl$_2(7\pi/6)=-\mbox{Cl}_2(5\pi/6)$,
Cl$_2(11\pi/6)=-\mbox{Cl}_2(\pi/6)$, $(1/2)\mbox{Cl}_2(\pi/3)=\mbox{Cl}_2
(\pi/6)-\mbox{Cl}_2(5\pi/6)$, Cl$_2(\pi/6)=(2/3)G+(1/4)\mbox{Cl}_2(\pi/3)$, and 
Cl$_2(5\pi/6)=(4/3)G - \mbox{Cl}_2(\pi/6)$ have been used.
In addition, we find
$$\sum_{n=1}^\infty {1 \over n^2}\left[\gamma+\psi\left(4n+{1 \over 2}\right)
\right ]=56\zeta(3)-4\pi G-16\pi \mbox{Cl}_2(\pi/4)-2\ln 2 \zeta(2), 
\eqno(14)$$
and
$$\sum_{n=1}^\infty {{(-1)^n} \over n^2}\left[\gamma+\psi\left(4n+{1 \over 2}
\right)\right ]=56\zeta(3)+7\pi \mbox{Cl}_2(\pi/4)-2\pi G-16\pi \mbox{Cl}_2
(3\pi/8)$$
$$-2\pi\mbox{Cl}_2(7\pi/8)+\zeta(2)\ln 2-14\pi\mbox{Cl}_2(\pi/8).
\eqno(15)$$
The details of deriving Eqs. (14) and (15) are omitted, although we note
that in the process we determined that
$\sum_{n=1}^\infty {{(-1)^n} \over n^2}\left[\gamma+\psi\left(4n\right)\right]
=(253/32)\zeta(3)+(\pi/2)G-4\pi \mbox{Cl}_2(\pi/4)$.

Similar to the proof of Eqs. (3), we have the following 
{\newline \bf Proposition}
$$\sum_{n=1}^\infty {1 \over n^2}\left[\gamma+\psi\left(kn+{1 \over 2}\right)
\right]={7 \over 2}k^2 \zeta(3)+2\pi\sum_{j=1}^{2k-1} j\mbox{Cl}_2(\pi j/k)
-\pi \sum_{j=1}^{k-1} j\mbox{Cl}_2(2\pi j/k)-2 \zeta(2) \ln 2, \eqno(16a)$$
$$\sum_{n=1}^\infty {{(-1)^n} \over n^2}\left[\gamma+\psi\left(kn+{1 \over 2}
\right)\right] = {7 \over 2}k^2 \zeta(3)+2\pi\sum_{j=1}^{2k-1} j\mbox{Cl}_2
\left({{\pi j} \over k} + {\pi \over {2k}}\right)$$
$$-\pi \sum_{j=1}^{k-1} j\mbox{Cl}_2\left({{2\pi j} \over k}+{\pi \over k}
\right)+\zeta(2) \ln 2, \eqno(16b)$$
In evaluating such sums, the Clausen function at rational multiples of $\pi$ 
can always be written in terms of the trigamma function at rational argument
and the sine function at rational multiples of $\pi$.  Specifically, we
have for integers $q \geq 2$ \cite{grosjean},
$$\mbox{Cl}_2\left({p \over q}\pi\right)={1 \over {4q^2}}\sum_{r=1}^{q-1}
\left[\psi'\left({r \over {2q}}\right)-\psi'\left(1-{r \over {2q}}\right)
\right]\sin r{p \over q}\pi$$
$$={1 \over {2q^2}}\sum_{r=1}^{q-1} \psi'\left({r \over {2q}}\right)\sin
\left(r {p \over q}\pi\right)-{\pi^2 \over {4q^2}}\sum_{r=1}^{q-1}
{{\sin(rp\pi/q)} \over {\sin^2(r\pi/2q)}}.  \eqno(17)$$
(By the use of the discrete sine transform, Eq.\ (17) can be inverted to
yield values of the trigamma function in terms of the sine and Clausen
functions.)
%later:  add in formula here .............

\medskip
\centerline{\bf Extension to sums over $(\pm 1)^n \psi(n+p/q)/n$}
\medskip

Parallel to the derivation of Eq. (6) from Eq. (5), we have
$$\sum_{n=1}^\infty {1 \over n^p}\sum_{r=0}^{q-1} \left[\gamma+\psi\left(n+
{r \over q}\right)\right ]=q\sum_{n=1}^\infty {1 \over n^p}\left[\gamma
+\psi\left(qn\right)\right ]-q\ln q\zeta(p), ~~\mbox{Re} ~p > 1, \eqno(18a)$$
and
$$\sum_{n=1}^\infty {{(-1)^n} \over n^p}\sum_{r=0}^{q-1} \left[\gamma+\psi
\left(n+{r \over q}\right)\right ] = q\sum_{n=1}^\infty {{(-1)^n} \over n^p}
\left[\gamma+\psi\left(qn\right)\right ]+q \ln q(1-2^{1-p})\zeta(p),
~~\mbox{Re} ~p > 1.  \eqno(18b)$$
When $p \to 1$, Eq. (18a) diverges, while Eq. (18b) is replaced by
$$\sum_{n=1}^\infty {{(-1)^n} \over n}\sum_{r=0}^{q-1} \left[\gamma+\psi
\left(n+{r \over q}\right)\right ] = q\sum_{n=1}^\infty {{(-1)^n} \over n}
\left[\gamma+\psi\left(qn\right)\right ]+q \ln q \ln 2. \eqno(19)$$
By using a result of Appendix D of Ref. \cite{ogreid98}, we obtain
$$\sum_{n=1}^\infty {{(-1)^n} \over n}\sum_{r=0}^{q-1} \left[\gamma+\psi
\left(n+{r \over q}\right)\right ] = q\left \{{1 \over 4}\left({1 \over q}
-q\right )\zeta(2)+{1 \over 2}\sum_{j=1}^q \left[\mbox{Cl}_1\left({{2\pi j}
\over q}+{\pi \over q}\right)\right]^2\right \}+q \ln q \ln2, \eqno(20)$$
where
$$\mbox{Cl}_1(\theta)=\sum_{k=1}^\infty {{\cos k\theta} \over k}
=-\ln |2\sin(\theta/2)|.  \eqno(21)$$
The $q=0$ and $q=1$ cases of Eq. (20) lead to
$$\sum_{n=1}^\infty {{(-1)^n} \over n}\left[\gamma+\psi
\left(n \right)\right ] = {1 \over 2}\ln^2 2, \eqno(22a)$$
and
$$\sum_{n=1}^\infty {{(-1)^n} \over n}\left[\gamma+\psi
\left(n +{1 \over 2} \right)\right ] = -{3 \over 4}\zeta(2)+2\ln^2 2. 
\eqno(22b)$$

\medskip
\centerline{\bf Generalization to sums over $(\pm 1)^n \psi(kn)/n^p$}
\medskip

The generalization to linear Euler sums with higher inverse powers of
$n$ is of continuing interest to both applications and theory.  Accordingly
we present new integral representations and show that special cases thereof
recover known results.  We have
{\newline \bf Proposition}
$$\sum_{n=1}^\infty {{(\pm 1)^n} \over n^p}[\gamma+\psi(kn+1)]={{\pm 1} \over 
{\Gamma(p)}} \int_0^\infty du {{u^{p-1} e^u} \over {e^u \mp 1}}\int_0^1
\left({{1-t^k} \over {1-t}}\right) {{dt} \over {(e^u \mp t^k)}},
~~~~\mbox{Re} ~p > 1.  \eqno(23)$$
In the case of the alternating sum, Eq. (23) holds for Re $p \geq 1$.
Furthermore, when $k$ is an integer, this equation may be rewritten as
$$\sum_{n=1}^\infty {{(\pm 1)^n} \over n^p}[\gamma+\psi(kn+1)]={{\pm 1} \over 
{\Gamma(p)}} \int_0^\infty du {{u^{p-1}} \over {1 \mp e^{-u}}}\sum_{j=0}^{k-1}
\int_0^1 {{t^j dt} \over {e^u \mp t^k}}.  \eqno(24)$$
With the aid of the functional equation for the digamma function, one has
the relation
$$\sum_{n=1}^\infty {{(\pm 1)^n} \over n^p}[\gamma+\psi(kn+1)]= 
\sum_{n=1}^\infty {{(\pm 1)^n} \over n^p}[\gamma+\psi(kn)] + {1 \over k}
\left \{\begin{array}{l}
\zeta(p+1) ~~~~\mbox{"top" sign}\\
(2^{-p}-1)\zeta(p+1) ~~~~\mbox{"bottom" sign}
\end{array} \right.
\eqno(25)$$
With the change of variable $v=t^k$, the integral over $t$ in Eq. (24)
can be written in terms of the hypergeometric function \cite{andrews},
%does this then already connect up with O & O's other method?
$$\int_0^1 {{t^j dt} \over {e^u \mp t^k}}= {e^{-u} \over {j+1}} ~_2F_1\left(1,
{{1+j} \over k};{{1+j+k} \over k}; \mp e^{-u}\right ). \eqno(26)$$
Then changing variable to $x=\exp(-u)$, another form of Eq.\ (24) is
%[look at similar steps for Eq. (23) too ....]
$$\sum_{n=1}^\infty {{(\pm 1)^n} \over n^p}[\gamma+\psi(kn+1)]={{\pm 1 (-1)^{p-1}} 
\over {\Gamma(p)}} \sum_{j=0}^{k-1} {1 \over {j+1}}\int_0^1 {{\ln^{p-1}} x
\over {1 \mp x}} ~_2F_1\left(1,{{1+j} \over k};{{1+j+k} \over k}; \mp x \right 
) dx. \eqno(27)$$

Equations (23) and (24) show the special character of the sum when $k=1$.
In this case, the integral over $t$ is elementary, and a change of variable
gives the representation
$$\sum_{n=1}^\infty {{(\pm 1)^n} \over n^p}[\gamma+\psi(n+1)]={{(-1)^p} \over 
{\Gamma(p)}} \int_0^1 {{\ln^{p-1} v} \over {v(1 \mp v)}} \ln (1 \mp v) dv
\eqno(28)$$
$$={{(-1)^p} \over {\Gamma(p)}} \int_0^1 \left ({1 \over v}+{1 \over {1-v}}
\right ) \ln^{p-1} v \ln (1 - v) dv, ~~~~\mbox{"top" sign}$$
$$={{(-1)^p} \over {\Gamma(p)}} \int_0^1 \left ({1 \over v}-{1 \over {1+v}}
\right ) \ln^{p-1} v \ln (1 + v) dv, ~~~~\mbox{"bottom" sign}.$$
The subcase at $n=3$ recovers results equivalent to those of de Doelder 
\cite{doelder} and Sitaramachandra Rao \cite{rao}:
%does he have an incorrect coefficient for the $\pi^4$ term ??? ... to look at ...
%--for my equ. (27) and also the ln^2 2 term ?? ...
$$\sum_{n=1}^\infty {1 \over n^3}[\gamma+\psi(n+1)]={5 \over 4}\zeta(4), 
\eqno(29)$$
$$\sum_{n=1}^\infty {{(-1)^n} \over n^3}[\gamma+\psi(n+1)]=
-{{11} \over 4}\zeta(4)+2\mbox{Li}_4(1/2)+{7 \over 4}\zeta(3)\ln 2 - {\pi^2 
\over {12}} \ln^2 2 + {1 \over {12}}\ln^4 2, \eqno(30)$$
with Eq. (30) exhibiting a needed correction to Eq. (8) of Ref.\ \cite{doelder}.
We also obtain %give more details ...
$$\sum_{n=1}^\infty {1 \over n^4}[\gamma+\psi(n+1)]=3 \zeta(5)- \zeta(2) 
\zeta(3).  \eqno(31)$$
Equation (30) shows the needed correction in de Doelder's Eq.\ (8) in both
the $(\pi^2/12) \ln^2 2$ and $-11\pi^4/360$ terms.  The corresponding 
corrected integral is given by
$$\int_{1/2}^1 {{\ln^2 u \ln (1-u)} \over u} du = 2\mbox{Li}_4(1/2)- {\pi^4 
\over {45}} + {7 \over 4}\zeta(3) \ln 2-{\pi^2 \over {12}}\ln^2 2 - {1 \over 6}
\ln^4 2.  \eqno(32)$$
%how about the other integral over 0 to 1/2 ?--is ok

For the case of $p$ an odd integer in the alternating sum of Eq. (28), 
Flajolet and Salvy \cite{flajolet} introduced the constants $\mu_q$,
with Eq. (30) giving $-\mu_1$.  Here the explicit form of these constants
emerges from our novel integral representations (23) and (28).
%[can I always evaluate part of Eq. (28) for general $p$ for the 
%alternating sum ????????]

For the proof of Eq. (23), we first use an integral representation of
the digamma function, giving
$$\sum_{n=1}^\infty {{(-1)^n} \over n^p}[\gamma+\psi(kn+1)]=
\sum_{n=1}^\infty {{(\pm 1)^n} \over {n^p}}\int_0^1 \left ({{1-t^{kn}} \over
{1-t}} \right ) dt.  \eqno(33)$$
We then introduce the representation $n^{-p}=[\Gamma(p)]^{-1}\int_0^\infty
e^{-nu} u^{p-1} du$ and interchange the summation and the integrations.
The needed sum is a simply a geometric series, and Eq. (23) follows.

\medskip
\centerline{\bf Extension to polygamma series}
\medskip

We now discuss the extension to polygamma series.  We have
{\newline \bf Proposition}
$$\sum_{n=1}^\infty {{(\pm 1)^n} \over n^p}\psi^{(j)}(kn+1)
={{\pm 1} \over {\Gamma(p)}}\int_0^\infty u^{p-1} e^{-u} du
\int_0^1 {{t^k \ln^j t dt} \over {(1-t)(1 \mp t^k e^{-u})}}, 
~~~~\mbox{Re} ~ p + j > 1.  \eqno(34)$$ 
%[check p range; is .=0 for the alternating sum,....?]  
For the alternating sum, the range of convergence extends to Re $p + j >0$.
With a change of variable, we have the alternative representation
$$\sum_{n=1}^\infty {{(\pm 1)^n} \over n^p}\psi^{(j)}(kn+1)
={{\pm 1(-1)^p} \over {\Gamma(p)}}\int_0^1 \ln^{p-1}v dv
\int_0^1 {{t^k \ln^j t dt} \over {(1-t)(1 \mp t^k v)}}, 
~~~~\mbox{Re} ~ p + j > 1.  \eqno(35)$$ 
By using the functional equation of the polygamma function \cite{nbs}, 
we have the relation
$$\sum_{n=1}^\infty {{(\pm 1)^n} \over n^p}\psi^{(j)}(kn+1)= 
\sum_{n=1}^\infty {{(\pm 1)^n} \over n^p}\psi^{(j)}(kn) + {{(-1)^j j!} \over 
k^j}\left \{\begin{array}{l}
\zeta(p+j+1) ~~~~\mbox{"top" sign}\\
(2^{-(p+j)}-1)\zeta(p+j+1) ~~~~\mbox{"bottom" sign}
\end{array} \right.
\eqno(36)$$

Equation (34) follows from
$$\sum_{n=1}^\infty {{(\pm 1)^n} \over n^p}\psi^{(j)}(kn+1)= 
-\sum_{n=1}^\infty {{(\pm 1)^n} \over n^p}\int_0^1 {t^{kn} \over {(1-t)}}
\ln^j t dt, \eqno(37)$$
introducing an integral representation for $n^{-p}$, and interchanging
summation and integration.

The very special case of $j=p=k=1$ gives
$$\sum_{n=1}^\infty {{(\pm 1)^n} \over n}\psi'(n+1)= 
\left \{\begin{array}{l}
\zeta(3)  \\
\zeta(3)-{\pi^2 \over 4}\ln 2
\end{array} \right.  \eqno(38)$$
%[later:  try doing the corresponding alternating sum for general $j$ .......]
The top line of Eq. (38) recovers a known result \cite{ogreid00}.  For the
alternating sum, we used \cite{devoto}
$$\int_0^1 {{\ln y \ln(1+y)} \over y}dy = 2S_{1,2}(-1)-\mbox{Li}_3(-1)+\ln 2
[\mbox{Li}_2(-1)-\zeta(2)], \eqno(39)$$
and $S_{1,2}(-1)=\zeta(3)/8$.

We also obtain
$$\sum_{n=1}^\infty (-1)^n \psi'(n)=-{3 \over 4}\zeta(2)=-{\pi^2 \over 8},
\eqno(40)$$
from \cite{devoto}
$$\int_0^1 {{\ln t ~dt} \over {1-t^2}}={1 \over 2}\int_0^1\left({1 \over
{1-t}} + {1 \over {1+t}}\right ) \ln t ~dt={1 \over 2}\left[\int_0^1 
{{\ln(1-y)} \over y}dy - {1 \over 2}\zeta(2)\right].  \eqno(41)$$

More generally, we have for integer $j$
$$\sum_{n=1}^\infty (-1)^n \psi^{(j)}(n)=\int_0^1 {{\ln^j t ~dt} \over 
{1-t^2}}={1 \over 2}\int_0^1 \left({1 \over
{1-t}} + {1 \over {1+t}}\right ) \ln^j t ~dt$$ 
$$= (-1)^j [2^{-(j+1)}-1]j!\zeta(j+1), \eqno(42)$$
and 
$$\sum_{n=1}^\infty {{(\pm 1)^n} \over n}\psi^{(j)}(n+1)= 
\int_0^1 {{\ln^j t \ln(1 \mp t)} \over {1-t}} dt
=\int_0^1 {{\ln^j (1 \mp w) \ln w} \over w} dw$$
$$=\pm {j \over 2}\int_0^1 {{\ln^{j-1}(1 \mp w)} \over {1-w}} \ln^2 w ~dw, 
\eqno(43)$$
where in the last step we integrated by parts.  For Eq. (42) we combined
the even and odd cases for $j$ of Ref. \cite{grad}.
%[what's the case j=2 result, etc.   can the general explicit result be
%given ?? ..............................................................]
We may record a few special-case integrals for use in conjunction with Eq.\ 
(43):
$$\int_0^1 \ln(1+t) {{\ln^2 t ~dt} \over {1-t}}=2[1-\zeta(3)], \eqno(44a)$$
$$\int_0^1 \ln(1+t) {{\ln^3 t ~dt} \over {1-t}}={1 \over {15}}(\pi^4-90)
\ln 2, \eqno(44b)$$
and
$$\int_0^1 \ln(1+t) {{\ln^4 t ~dt} \over {1-t}}=-24\ln2[\zeta(5)-1]. 
\eqno(44c)$$

As an extension of Eqs. (34) and (35), we may present a five-parameter 
summation in:
{\newline \bf Proposition}
$$\sum_{n=1}^\infty {{(\pm 1)^n} \over n^p}\psi^{(j)}(kn + a) z^n
={{\pm z(-1)^p} \over {\Gamma(p)}}\int_0^1 \ln^{p-1}v dv
\int_0^1 {{t^{k+a-1} \ln^j t dt} \over {(1-t)(1 \mp zv t^k)}}, 
~~~~\mbox{Re} ~ p + j > 1.  \eqno(45)$$ 
On the left side of this equation, the $(\pm 1)^n$ factor could just as
well be absorbed in the new $z^n$ factor.  The proof of Eq. (45) proceeds
as for Eq. (34), again followed by a change of variable.  The details are
omitted.

\medskip
\centerline{\bf Additional digamma and polygamma series}
\medskip 

The many possible extensions of our work include digamma and polygamma
series with these functions at rational arguments and denominators of
the form $n^\alpha (n+1)^\beta (n+2)^\gamma \cdot\cdots$.  We present in
passing a few simple examples:
$$\sum_{n=1}^\infty {{[\gamma+\psi(n+1/2)]} \over {n(n+1)}}
=2 \ln 2, \eqno(46)$$
$$\sum_{n=1}^\infty {{[\gamma+\psi(n+1/2)]} \over {n^2(n+1)}}
={7 \over 2}\zeta(3)-2\ln 2-2\zeta(2)\ln 2, \eqno(47)$$
and
$$\sum_{n=1}^\infty {{[\gamma+\psi(n+1/2)]} \over {n(n+1)^2}}
=-{7 \over 2}\zeta(3)-2\zeta(2)+8\ln 2+2\zeta(2)\ln2. \eqno(48)$$
%will Mca do these, and will it give the corresponding alternating series??
As an area of possible future research, these sums suggest the
investigation of series of the form
$$\sum_{n=1}^\infty (\pm 1)^n{{[\gamma+\psi(n+p/q)]} \over {{n \choose j}^\alpha
{n \choose \ell}^\beta \cdots}} \eqno(49a)$$
and
$$\sum_{n=1}^\infty (\pm 1)^n {{\psi^{(m)}(n+p/q)} \over {{n \choose j}^\alpha
{n \choose \ell}^\beta \cdots}} \eqno(49b)$$
where $m$, $p$, $q$, $j$, $\ell$, ... and $\alpha$, $\beta$, ... are integers.
%completely outside the scope here would then be the nonlinear extension
%--powers of the digamma and polygamma functions ...

By using the series
$$\sum_{\ell=1}^\infty {t^\ell \over {\ell(\ell+1)(\ell+2)\cdots(\ell+j)}}
={t \over {(j+1)!}} ~_2F_1(1,1;j+2;t), \eqno(50)$$
and an integral representation for the digamma function, we find
{\newline \bf Proposition} for integers $j \geq 1$
$$\sum_{n=1}^\infty {{[\gamma+\psi(n+1)]} \over {n(n+1)\cdots(n+j)}}
={1 \over {(j+1)!}}\int_0^1 \left [1 + {1 \over j} - t ~_2F_1(1,1;j+2;t)
\right ] {{dt} \over {1-t}}.  \eqno(51)$$
From the series in Eq. (50), we can multiply by $t^j$ and integrate from
$0$ to $z$ to find
$$\sum_{\ell=1}^\infty {z^\ell \over {\ell(\ell+1)(\ell+2)\cdots(\ell+j)
(\ell+j+1)}}={1 \over z^{j+1}}\int_0^z t^j f(t)dt, \eqno(52)$$
where $f(t)$ denotes the right side of Eq. (50).  The point here is that
we have, for instance, $[t+(1-t)\ln(1-t)]/t=(t/2) ~_2F_1(1,1;3;t)$ so that
we can always successively obtain the needed hypergeometric functions in
Eq. (51) in terms of logarithmic functions.

Generalizations of Eq. (50) are readily performed.  We have
{\newline \bf Proposition} for integers $j \geq 1$
$$\sum_{n=1}^\infty {{[\gamma+\psi(kn+1)]} \over {n(n+1)\cdots(n+j)}}
={1 \over {(j+1)!}}\int_0^1 \left [1 + {1 \over j} - t^k ~_2F_1(1,1;j+2;t^k)
\right ] {{dt} \over {1-t}}, \eqno(53)$$
and
$$\sum_{n=1}^\infty {{\psi^{(m)}(kn+1)} \over {n(n+1)\cdots(n+j)}}
=-{1 \over {(j+1)!}}\int_0^1 t^k ~_2F_1(1,1;j+2;t^k) {{\ln^m t dt} 
\over {1-t}}.  \eqno(54)$$
Just as we derived Eqs. (3) and (18) from the multiplication formula for
the digamma function, Eqs. (51)-(54) can serve to construct generalizations
of Eq. (46) and related sums.

As a set of auxiliary sums complementing the previous sums of this section,
we present the following
{\newline \bf Proposition} for integers $j \geq 1$
$$\sum_{n=1}^\infty {{(-1)^n} \over {n(n+1)^{j+1}}}=\sum_{m=1}^j (2^{-m}-1)
\zeta(m+1) + j + 1 - 2\ln 2.  \eqno(55)$$ 
By the use of partial fractions, these sums lead to several related ones.
Since these are alternating sums with decreasing summand and first term 
negative, their values are negative for all $j$.

For the proof of Eq. (55), we first derive
$$\sum_{n=1}^\infty {{(-1)^n} \over {n(n+z)}}={1 \over z}\left[\psi\left(1+
{z \over 2}\right)-\psi\left(1+z\right)\right].  \eqno(56)$$
We then differentiate this equation $j$ times to find
$$\sum_{n=1}^\infty {{(-1)^n} \over {n(n+z)^{j+1}}}=\sum_{m=0}^j {{(-1)^m}
\over {m!}} {1 \over z^{j-m+1}} \left[{1 \over 2^m} \psi^{(m)}\left(1+
{z \over 2}\right) - \psi^{(m)} \left(1+z\right)\right].  \eqno(57)$$
We evaluate Eq. (56) at $z=1$, separating the $m=0$ term of the sum:
$$\sum_{n=1}^\infty {{(-1)^n} \over {n(n+1)^{j+1}}}=\sum_{m=1}^j {{(-1)^m}
\over {m!}} \left[2^{-m} \psi^{(m)}\left(
{3 \over 2}\right) - \psi^{(m)} \left(2\right)\right]
+ \psi\left({3 \over 2}\right)-\psi(2).  \eqno(58)$$
To find Eq. (55), we then use the special values $\psi(2)=1-\gamma$, 
$\psi(3/2)=2-\gamma-2\ln 2$, $\psi^{(j)}(2)=(-1)^{j+1}j![\zeta(j+1)-1]$,
and $\psi^{(m)}(3/2)=(-1)^{m+1}m!\{2^{m+1}[\zeta(m+1)-1]-\zeta(m+1)\}$.

We next indicate an alternative proof of Eq. (55).  This method of proof
works equally well in determining 
{\newline \bf Proposition}
$$\sum_{n=1}^\infty {{(-1)^n} \over {n^{j+1}(n+1)}}=(-1)^j \left[\sum_{m=1}^j 
(-1)^m(2^{-m}-1) \zeta(m+1) + 1 - 2\ln 2 \right].  \eqno(59)$$ 
In the case of Eq. (55), we put
$$T_j \equiv \sum_{n=1}^\infty {{(-1)^n} \over {n(n+1)^{j+1}}}
=\sum_{n=1}^\infty {{(-1)^n} \over {(n+1)^j}}\left({1 \over n}-{1 \over {n+1}}
\right).  \eqno(60)$$
Therefore, we easily find the recursion relation
$T_j=T_{j-1}+(2^{-j}-1)\zeta(j+1)+1$, with $T_0=1-2\ln 2$.  Iteration of
this recursion relation gives Eq. (55).

Similarly, if we put
$$S_j \equiv \sum_{n=1}^\infty {{(-1)^n} \over {n^{j+1}(n+1)}}
=\sum_{n=1}^\infty {{(-1)^n} \over {n^j}}\left({1 \over n}-{1 \over {n+1}}
\right),  \eqno(61)$$
we quickly find the recursion relation
$S_j = -S_{j-1}+(2^{-j}-1)\zeta(j+1)$, with $S_0=T_0$.  Iteration of
this recursion relation leads to Eq. (59).

By using the functional equations of the digamma and polygamma functions,
Eq. (57) can be rewritten as 
$$\sum_{n=1}^\infty {{(-1)^n} \over {n(n+z)^{j+1}}}=\sum_{m=1}^j {{(-1)^m}
\over {m!}} {1 \over z^{j-m+1}} \left[2^{-m} \psi^{(m)}\left({z \over 2}
\right) +(-1)^m2^{-m}m!\left({z \over 2}\right)^{-m-1}\right.$$
$$\left. - \psi^{(m)} \left(z \right)-(-1)^mm!z^{-m-1}\right]
+{1 \over z^{j+1}}\left[\psi\left({z \over 2}
\right)-\psi(z)+{1 \over z}\right].  \eqno(62)$$
Evaluation of this equation at rational values of $z$ offers a multitude of
other sums.  We have $\psi^{(n)}(x)=(-1)^{n+1}n!\zeta(n+1,x)$ \cite{nbs},
in terms of the Hurwitz zeta function $\zeta(s,a)$, and evaluation of the
Hurwitz zeta function at rational argument presents no difficulty.  For
instance, we have $\psi^{(m)}(1/4)=(-1)^{m+1}m!\zeta(m+1,1/4)$ for use in
evaluating the sum $\sum_{n=1}^\infty (-1)^n/[n(n+1/2)^{j+1}]$.  By combined
use of the reflection and duplication formulas for the digamma function, we
have simply $\psi(1/4)=-\gamma-3\ln 2-\pi/2$.  %value to check in M'ca ...
%give kulovic and c expression in here ???

As a further development, we present
{\newline \bf Proposition}  For integers $N \geq 1$
$$\sum_{\ell=1}^\infty {1 \over \ell}{1 \over {\prod_{j=0}^N (\ell+j)}}
=\zeta(2) + \sum_{k=1}^N {{(-1)^k} \over k} {N \choose k} [\psi(k+1)+\gamma],
\eqno(63a)$$
$$\sum_{\ell=1}^\infty {{(-1)^\ell} \over \ell}{1 \over {\prod_{j=0}^N (\ell
+j)}} = -{1 \over 2}\zeta(2) + \sum_{k=1}^N {{(-1)^k} \over k} {N \choose k} 
\left[\psi\left(1+{k \over 2}\right) -\psi(1+k)\right], \eqno(63b)$$
The $N=1$ case of Eqs. (63) gives the above $S_1$ and $T_1$ sums.
The proof of Eqs. (63) proceeds from the well known expansion
$$\sum_{k=0}^N {N \choose k} {{(-1)^k} \over {x+k}}={N \over {\prod_{j=0}^N
(x+j)}}.  \eqno(64)$$
We then multiply both sides of this equation by $(\pm 1)^x/x$ and sum over
$x$ from $1$ to $\infty$.  For Eq. (63b), Eq. (56) is applied, while for
Eq. (63a), the widely known summation
$\sum_{n=1}^\infty [n(n+z)]^{-1}=[\psi(1+z)+\gamma]/z$ is applied.
The $k=0$ term is treated by noting the expansion \cite{grad}
$$\psi(1+x)=-\gamma+\sum_{k=2}^\infty (-1)^k \zeta(k) x^{k-1}.  \eqno(65)$$
This gives, for instance, $\lim_{k \to 0} [\psi(1+k)+\gamma]/k = \zeta(2)$,
and leads to Eqs. (63).

The determination of the nonalternating sums corresponding to Eqs. (55) and
(59) is easily made.  We find
$$\sum_{n=1}^\infty {1 \over {n^{j+1}(n+1)}}=(-1)^j\left[\sum_{m=1}^j (-1)^m
\zeta(m+1) + 1 \right ], \eqno(66a)$$
and
$$\sum_{n=1}^\infty {1 \over {n(n+1)^{j+1}}}=-\sum_{m=1}^j \zeta(m+1)+j+1.
\eqno(66b)$$
These sums may be derived either from a recursion relation or from a summation
in terms of polygamma functions. 

The sums of this section are intimately linked to hypergeometric summation.
For instance, we have $_3F_2(1,1,1;2,3;t)=2\sum_{k=1}^\infty t^{k-1}/k^2(k+1)$
and more generally
$$_{m+1}F_m(1,1,\ldots,1;2,2,\ldots,2,3;t)=2\sum_{k=0}^\infty {t^{k-1} \over
{k^m(k+1)}}.  \eqno(67)$$
Of course, the polylogarithm itself is given by Li$_m(x)=x ~_{m+1}F_m(1,1,
\ldots,1;2,\ldots,2;x)$.  Particularly in cases of the $_3F_2$ function
evaluated at unit argument, the reductions of Watson, Whipple, and Dixon
\cite{andrews,weng} can be brought to bear.  However, we do not pursue this 
direction here.

%somewhere to present comparison integral(s) (estimations for large $k$, ...)
%+ remarks related to sums from hypergeometric series ....

%\medskip
\pagebreak
\centerline{\bf Summary}
\medskip

Among many developments, we have investigated summations over digamma and 
polygamma functions at rational argument.  We have considered both positive
and alternating series.  By exploiting properties of the Clausen function
Cl$_2$ we have presented explicit, simplified expressions in many cases.
We have developed novel general integral representations and given explicit
instances.  Special cases of the sums reduce to known linear Euler sums, or
to related digamma and polygamma series.  
We have mentioned connections to the theory of polylogarithms, 
hypergeometric functions, and zeta functions.

%The sums of interest find application in quantum field theory, including
%evaluation of Feynman amplitudes.  %NEED TO REWORD/REPLACE THAT SENTENCE ...
The sums of interest have relevance to calculations in quantum electrodynamics.
In particular, they support the determination of Feynman amplitudes.
Furthermore, our sums appear at a sort of
cross roads of mathematical physics and analytic number theory.  Part of the
relevance to analytic number theory is the connection of subcases to
Euler sums, and representations in terms of the Riemann and Hurwitz zeta
functions.  In addition, our results
may serve as a type of test bed for symbolic computation, and as a stimulus
for investigation of further linear sums as in Eqs. (49).

\pagebreak
\centerline{\bf Appendix:  Simple sum estimations}
\medskip

In the sum
$$\sum_{n=1}^\infty (\pm 1)^n {{[\gamma+\psi(kn+1)]} \over n^p}
=\sum_{n=1}^\infty  \sum_{j=1}^{kn} {{(-1)^n} \over {n^p j}}, \eqno(A.1)$$
where $\psi=\Gamma'/\Gamma$ is the digamma function, it is well known that
$$\sum_{j=1}^{kn} {1 \over j}=\ln (kn)+\gamma+o(1), \eqno(A.2)$$
where $\gamma$ is the Euler constant.
Therefore, for the sum
$$\sum_{n=1}^\infty {{[\gamma+\psi(kn+1)]} \over n^p}=\sum_{n=1}^\infty
{{\ln n} \over n^p}+\sum_{n=1}^\infty {{\ln k+\gamma +o(1)} \over n^p},
\eqno(A.3)$$
the leading behaviour is given by the comparison integral
$\int_1^\infty (\ln x)/x^p dx=\Gamma(2)/(p-1)^2$.  Therefore there is absolute
convergence of the sum of (A.1) for Re $p >1$.  On the other hand, the
alternating sum in Eq. (A.1) converges for Re $p >0$.

Similarly, for polygamma series we can make use of the known asymptotic
expansion \cite{nbs}.  As a very simple example, the leading behaviour
of $\sum_{n=1}^\infty \psi^{(m)}(kn)/n^p$ is given by $\int_1^\infty
x^{-p-m} dx$.  Therefore absolute convergence obtains when Re$(p + m) >1$.

%\pagebreak
%\bigskip
%\centerline{\bf Figure Caption}
%\medskip

%FIG. 1.  Caption would go here ...

\pagebreak

\end{document}